\documentstyle[psfig,conf_iap,]{article}
\begin{document}
\heading{
Small scale structure in diffuse molecular gas\\ 
from repeated FUSE and visible spectra of HD 34078
} 
%
\author{%
Patrick Boiss\'e$^{1}$, Emmanuel Rollinde$^{2}$, Franck Le
Petit$^{3}$, Guillaume Pineau des For\^ets$^3$, 
Evelyne Roueff$^3$, C\'ecile Gry$^{4}$, B-G Andersson$^{5}$,
 Vincent Le Brun$^4$.
}
\address{
Ecole Normale Sup\'erieure/Observatoire de Paris, France }
\address{
I.A.P., 98 bis Bd Arago, Paris
}
\address{
DAEC, Observatoire de Paris-Meudon
}
\address{
Laboratoire d'Astronomie Spatiale, Marseille, France
}
\address{
John Hopkins University, Baltimore, USA
}

\vskip -0.3cm
\begin{abstract}
We present preliminary results from an ongoing program devoted to a study of 
small scale structure in the spatial distribution of molecular gas. 
Our work is based on multi-epoch FUSE and visible observations of
HD34078. A detailed comparison of H$_2$, CH and 
CH$^+$ absorption lines is performed. 
No short term variations are seen (except for 
highly excited H$_2$) but long-term changes in 
N(CH) are clearly detected when comparing our data to spectra taken
about 10 years ago.
\end{abstract}

\section{Introduction}
The small scale distribution of molecular species in translucent and
opaque clouds is still poorly characterized as compared to structure
in the atomic phase (see the paper by J. Lauroesch in these
proceedings). 
To complement the scarce data available on minor tracers like H$_2$CO, 
OH, HCO$^+$ (\cite{mar93}, \cite{liszt}), we have undertaken an 
observational program devoted to a simultaneous 
study of small scale variations in the column density of various 
molecular and atomic species, including the major ones, H$_2$ and HI.

A bright, large proper motion star, HD 34078, has been
selected for repeated observations in the far-UV with FUSE, and in 
the visible range, at the Observatoire 
de Haute-Provence. Due to the large transverse velocity 
of this target (100km/s), the line of sight is drifting through the foreground 
material (E(B-V) = 0.52) at a velocity of about 17 AU/yr. Thus, observations 
separated by an interval ranging from about one month to a few years 
provide information on structure at scales 1 - 50 AU. 
Further, HD 34078 has been the subject of detailed studies in the past 
and several high S/N and resolution spectra have been obtained
(\cite{allen}, \cite{fed94}, \cite{herbig}). 
By comparison with our recent ones, scales as large as 200 AU can 
then be probed for species like CH and CH$^+$. 

The main properties of the absorbing material along the sightline
towards HD 34078 are discussed in Le Petit et al. (these proceedings). 
In this paper, we focus on absorption line $\it {variations}$ and 
present a preliminary analysis of H$_2$ 
features in the three FUSE cycle 1 spectra and of the
variations of visible CH and CH$^+$ absorption lines. 

\section{Observations and data analysis}
The FUSE spectra were obtained in january 2000, 
october 2000 and february 2001. The data have been 
processed using the FUSE pipeline software. Individual 
exposures were co-added in a standard manner. 
 For each epoch, the 
total integration time is about 6500s; the S/N ratio reached in the 
LiF1A data that we will use here is about 30 per 15 m\AA\ pixel. 
Seven visible spectra were obtained between 
november 1999 and february 2001 at the 
Observatoire de Haute Provence (O.H.P.) with the echelle spectrometer, ELODIE, 
mounted on the 1.93m telescope. The spectral resolution is about 10km/s; 
the typical S/N ratio is 150 per .03 \AA\ pixel. A spacing of about one month was 
adopted first but since CH and CH$^+$ lines looked quite stable, 
only two spectra were taken next winter. 
 
In FUSE spectra, a large number of transitions are available 
to study column density variations but in fact, only a few of 
them are suitable for a sensitive search for variations. 
Indeed, since profiles are not resolved by FUSE, the equivalent 
width of lines falling on the flat part of the curve of growth are 
insensitive to N variations. The situation is favorable in 
two cases: a) marginally thin lines (0.5 $< \tau_{0} <$ 3 where 
$\tau_{0}$ is the opacity at line center) and b) 
damped lines.  For species which are likely to display the same b 
value as CH (about 3 km/s), case a) corresponds to lines with W ranging 
from roughly 10 to 50 m\AA. H$_2$ lines from high excitation levels, 
CH and CH$^+$ lines fall into this category. 
Case b) corresponds to Ly$\beta$ or the 
strongest J = 0, 1, 2 H$_2$ lines. It can be shown easily that 
the difference between two damped profiles is  
maximum where the normalised intensity equals 0.36; 
typically, for a 10\% N variation, the 
expected difference is 0.04. Thus, a definite signature is
expected for such lines, which should be easily detectable in 
good quality spectra if $\Delta$N is large enough.
Obviously, some care has to be taken concerning 
instrumental effects which might induce apparent line variations but
using several transitions 
from the same species or data from different FUSE segments can 
be used to solve these potential difficulties.
\vskip -0.8cm
\section{Results}
\begin{figure}
\centerline{\vbox{
\psfig{figure=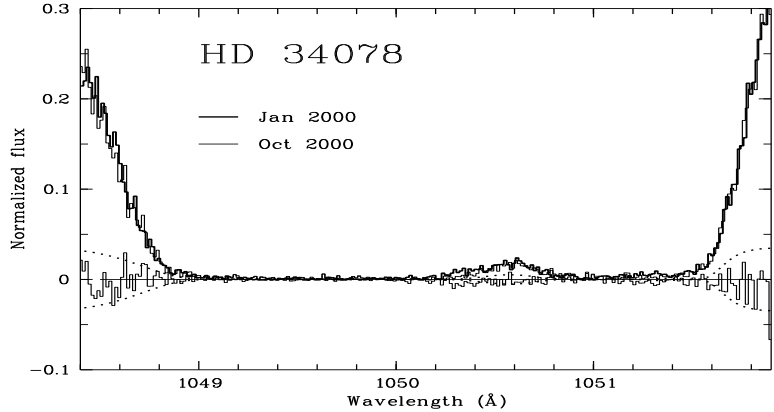,width=11cm,height=6cm,angle=-90}
}}
\caption[]{Extract of the FUSE Jan. 2000 and Oct. 2000 spectra
showing damped H$_2$ profiles and their difference. 
The dotted lines correspond to a $\pm$ 10\% variation in 
N(H$_2$, J=0,1 and 2)}.
\label{fig1}
\end{figure}

{\bf H$_2$:} the first goal of our project is to estimate the fluctuations 
of the total H$_2$ column density. We thus need to study lines from 
the J = 0, 1 and 2 states. Several damped lines originating from these 
levels and with good S/N profiles 
are present in our spectra. These features appear 
blended in systems but generally, the blue or red wing is primarily 
due to one specific transition (for instance, in the system around 
1050\AA\ shown in Fig. \ref{fig1}, the blue wing is due mostly to the J = 0 
transition); blends with other lines also 
restricts the range to be used. Before comparing profiles from 
distinct epochs, we first adjust the wavelength scales by applying a 
uniform shift. 
 Second, we adjust the flux level in intervals 
free of absorption using one single multiplicative factor. As can be
seen in Fig. \ref{fig1} profiles look extremely similar; their 
difference shows no significant deviation from zero. 
For comparison, we show in Fig. \ref{fig1} the difference expected if N(J = 0, 
1 and 2) are varied by $\pm$ 10\%. Our residuals are clearly 
well smaller ! We 
conclude that over the three epochs considered, a 2.5 $\sigma$ upper limit on 
variations in N(J=0 and J=2)  is 7\%. This result suggests that N(H$_2$) 
variations are smaller than those reported for H$_2$CO and OH.

For the highest excitation H$_2$ levels detected, lines become 
close to optically thin 
and we recover a good sensitivity to variations. 
We find some evidence for variations with W(oct. 
2000) $<$ W(feb. 2001) $<$ W(jan. 2000). For instance, in the 
region around 1046 \AA, large variations are seen for a feature at 
1046.45 \AA\ that we tentatively assign to H$_2$(v=4, J=2); 
similarly, two features from H$_2$ J=11 apparently show in the 
2d and 3rd spectra a blueshift of about 10 km/s and a broader
profile. By cross-correlating the 
behaviour of various lines from each level, we should be able to reach firm 
conclusions on the reality and magnitude of these variations. As discussed by 
Le Petit et al. in this volume, variable excited H$_2$ 
lines could be due to instabilities in the shocked gas at the 
HD34078's wind - ambient medium interface. 

{\bf CH and CH$^+$:} no small scale variations are detected in our homogeneous
set of OHP spectra. The excellent S/N reached on the CH4300 line
allows us to set a 3 $\sigma$ limit on relative 
column density fluctuations $\Delta$N/N $\leq$ 6\% for scales in the 
range 1 - 20AU. Thus, we do not find evidence for the kind of ubiquitous 
structure seen for OH and H$_2$CO \cite{mar93}.
\begin{figure}
\centerline{\vbox{
\psfig{figure=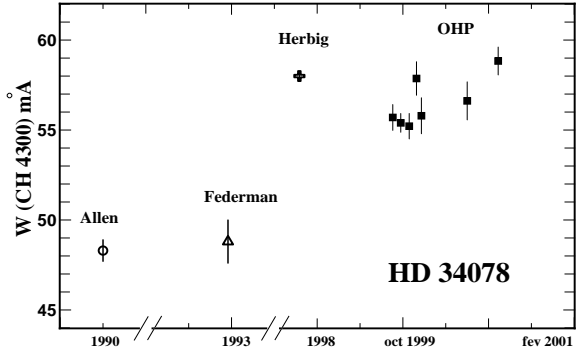,height=4.8cm,width=8cm,angle=90}
}}
\caption[]{Measurements for W(CH4300) as a function of time}
\label{fig2}
\end{figure}
However, comparing our results with those obtained several years 
ago (\cite{allen}, Federman, private communication, \cite{herbig}), 
we find that W(CH4300) has increased by about 18\% between 1993 and 
1998 as illustrated by Fig. \ref{fig2} (this interval corresponds 
to about 100 AU). It is noteworthy that, from the analysis of the 
damped Ly$\beta$ profile (blended with H$_2$ lines), we get an 
estimate of the HI column density, N(HI) = 2.51 10$^{21}$ cm$^{-2}$,
which is 38\% higher than the value derived from 1979 IUE data, 
N(HI) = 1.82 10$^{21}$ cm$^{-2}$ (\cite{maclac}; the time interval 
corresponds to 
about 300 AU).

Regarding CH$^+$, we get a less stringent 
limit on small scale variations $\Delta$N/N $\leq$ 15\% and 
do not find evidence for a similar large scale variation. 

Let us conclude with a few prospects. The coming cycle 2 and 3 FUSE 
spectra will clearly allow us to improve the sampling and to extend the 
range of scales probed. Thus, we will be more sensitive to a small 
regular variation (i.e. of N(H$_2$, J=0 or 1)). 
Additional spectra will also help to establish the reality of
variations suspected for the weak lines of highly excited H$_2$, 
and quantify the magnitude of these changes. Finally, we note 
that HST/STIS spectra would complement these FUSE data in a 
very useful way because a) the HST range comprises a lot of 
vibrationally excited H$_2$ transitions; b) given the spectral 
resolution of STIS, most 
of these lines would appear unblended; c) changes in the 
kinematics of the excited H$_2$ would be much more apparent. 

{\bf Acknowledgments:} We are grateful to Steve Federman for
communicating his 1993 W(CH4300) measurement and to Jim Lauroesch 
who kindly provided us with a co-added Copernicus $\zeta$ Oph spectrum.

\end{document}